# An Explainable XGBoost-based Approach on Assessing Detection of Deception and Disinformation

Alex V Mbaziira
College of Business, Innovation, Leadership and Technology
Marymount University, USA
ambaziir@marymount.edu

Maha F Sabir
Faculty of Computing & Information Technology
King Abdulaziz University, Saudi Arabia
msaber@kau.edu.sa

## Abstract

*Threat actors continue to exploit geopolitical and global public events launch aggressive campaigns propagating disinformation over the Internet. In this paper we extend our prior research in detecting disinformation using psycholinguistic and computational linguistic processes linked to deception and cybercrime to gain an understanding of the features impact the predictive outcome of machine learning models. In this paper we attempt to determine patterns of deception in disinformation in hybrid models trained on disinformation and scams, fake positive and negative online reviews, or fraud using the eXtreme Gradient Boosting machine learning algorithm. Four hybrid models are generated which are models trained on disinformation and fraud (DIS+EN), disinformation and scams (DIS+FB), disinformation and favorable fake reviews (DIS+POS) and disinformation and unfavorable fake reviews (DIS+NEG). The four hybrid models detected deception and disinformation with predictive accuracies ranging from 75% to 85%. The outcome of the models was evaluated with SHAP to determine the impact of the features.*

## 1. Introduction

Disinformation has continued to be a problem as threat actors exploit social media and Internet connected devices to disrupt democratic and civil processes through geopolitical events as well as delivery of public health services. For example, the US Presidential election of 2016 and global health crisis, are some of the infamous global events that exposed the disinformation problem and its impact on communities (Hollyer et al., 2019; Miller, 2020; Rocha et al., 2021). Disinformation can be propagated in different forms like audio, text and video, however, our paper is limited to investigating text-based disinformation where we focus on deception detection and explainable machine learning.

The models used in this paper are trained using the *eXtreme Gradient Boosting (XGBoost)* machine learning algorithm and *SHapley Additive exPlanations (SHAP)*, which is a popular explainable machine learning framework. SHAP has been widely adopted to provide insights on how features impact to the decision processes of models. The dataset for disinformation used is in this paper is fake news generated by trolls. This paper extends of some of previous papers on detecting deception in text-based cybercrime using computational linguistic and psycholinguistic processes linked to deception and cybercrime to determine on which this paper builds its foundation (A. Mbaziira & Jones, 2016; A. V. Mbaziira et al., 2022; A. V. Mbaziira & Jones, 2017; A. V. Mbaziira & Murphy, 2018). The research question which this paper investigates is: *How do computational linguistic and psycholinguistic features contribute to the decision process in detecting disinformation for hybrid models trained on fraud, scams, fake reviews*? The rest of the chapter is organized as follows: Section 2 presents the related work; Section 3 presents the methodology used in generating and evaluating our models. The results of the models and the analysis are presented in Section 4. Section 5 presents the conclusions for this research.

## 2. Related Work

In this section we briefly describe the SHAP framework and review of the related work on disinformation detection and adoption SHAP to explain results of the AI models. *SHapley Additive exPlanations* (*SHAP*) is one of the frameworks used in explain output of black box machine learning algorithms to ensure transparency, better decision making by explaining output of black box models, and model debugging. The main intuition behind Shapley value-based explanations of black-box models is use of fair allocation results from cooperative game theory to allocate credit for a model's output function f(x) within its input features (Lundberg & Lee, 2017). The framework generates *shapley* values which are then used to determine the payoffs of each player contributing to the game. Shapley values of each feature is calculated and it from these values the explanations of the models are generated. Below is the formula for generating *shapley* value for a given feature $x_0$:

$$\theta(x_0) = \sum_{S \subseteq N/i} \frac{(|S|!(n-|S|-1)!)}{n!} (x_0(SUi) - x_0(S)) \qquad (1)$$

One study investigated fake news campaign detection XGBoost machine learning algorithm to detect the similarity index in social media campaigns (Framewala et al., 2020). The study presents a novel semi-supervised learning approach for detecting fake news campaigns and giving human-like explanations for those detection. The study also uses a modified version of Artificial Immune System (AIS) for unsupervised detection of social media campaigns which include public outrage, marketing and public awareness.

Another study compares logistic regression and XGBoost algorithms in detecting counterfeit news (Rao et al., 2021). The paper uses term frequency and inverse document frequency (TF-IDF) to generate features used in generating models for fake news detection.

Another study on fake news detection using machine learning uses a SHAP for model explain-ability as well as psycholinguistics and n-grams for feature modelling (Villagracia Octaviano, 2021). The paper then explores various feature sets, whereby two new features are introduced to develop an automated fake news detector s. The models in the study for detecting fake news are generated from Linear Support Vector Machines, Convolutional Neural Networks, *XGBoost* and multi-layer perceptron.

The differences between our paper and the related work analyzed above are that in our case, we use SHAP to explain the results of our models and also use novel feature engineering approach. We use psycholinguistic and computation linguistic feature processes linked to deception and cybercrime to generate features which our models use to detect deception and disinformation.

### 3. Methodology

#### *a) Dataset Description*

The datasets used in training the models are the scam dataset was generated from public Facebook accounts on an actual cybercriminal network of scammers who were using an online data theft service (Sarvari et al., 2014). The Facebook accounts were obtained by performing lookups from an email dataset which was publicly leaked. Facebook data which comprised of scams used by this criminal network to exploit their victims were manually verified the scams and labelled. Benign data was collected from these accounts and also labelled. All scam data was labeled as deceptive while the benign data was labelled was truthful. For fraud, we considered two Enron corporation email datasets: the truthful dataset comprised of emails that were made public by the Federal Energy Regulatory Commission (Cohen, 2015) while the deceptive dataset comprised of evidence used in court to convict the Enron executives for securities and wire fraud. For fake online reviews, we used two public labelled datasets on favorable and unfavorable fake (Ott et al., 2011). For the disinformation dataset, we used a public fake news dataset (Verma et al., 2019). As shown in Table 1, 200 instances were randomly selected for generating the training set for each of the models while 20 examples were randomly selected to generate a test set of disinformation. Since we used hybrid models in each of the models, half of instances contained disinformation examples which were not part of the test set.

**Table 1: Dataset Description**

| Model | Dataset | Training Set (90%) | DIS Test set (10%) |
|---|---|---|---|
| DIS+EN | Disinformation & Enron | 200 | 20 |
| DIS+FB | Disinformation & Facebook | 200 | 20 |
| DIS+POS | Disinformation & Favorable Reviews | 200 | 20 |
| DIS+NEG | Disinformation & Unfavorable Reviews | 200 | 20 |

## b) Feature selection and engineering

We considered computational linguistic and psycholinguistic processes linked to deception and cybercrime when engineering features for our models. For purposes of this study, by cybercrime we referred to online committed in form fraud, scams and deceptive reviews because deception plays an important role in exploiting victims. These computational linguistic features which are linked to deception and cybercrime used in our models are: verbs, average sentence length, average word length, modal verbs, lexical diversity, characters, punctuation marks, sentences, adjectives, adverbs, nouns and function words (Zhou et al., 2004). Furthermore, we considered the following the features for psycholinguistic processes linked to deception and cybercrime also based on our earlier work: *analytical, 6-letter words, I and insight.* Table 2 is a summary of features used in the models.

**Table 2: Features used in the Models**

| Feature | Linguistic Process Linked to Deception and Cybercrime | Description of the Relevance of the Feature to Deception |
|---|---|---|
| Verbs | Computational Linguistic | Measures frequency of verbs whereby deceptive messages will have non-committal verbs than truthful messages |
| Modifiers | Computational Linguistic | Frequency of modifiers will be greater in deceptive messages compared to truthful messages |
| Averages sentence length | Computational Linguistic | Average sentence length of deceptive messages will be greater than in truth messages. |
| Average word length | Computational Linguistic | Deceptive messages will usually be shorter than truthful messages. |
| Modal verbs | Computational Linguistic | Frequency of verbs that indicate obligation. Since cybercriminals are non-committal, fewer modal verbs will be used in deceptive messages. |
| Lexical diversity | Computational Linguistic | Measure ratio of unique words per sentence. Deceptive messages will have lower lexical diversity compared to truthful messages. |

| Number of characters | Computational Linguistic | Frequency of characters in sentences. Deceptive messages will have higher character frequency compared to truthful messages. |
|---|---|---|
| Number punctuation marks | Computational Linguistic | Frequency of punctuation marks. Deceptive messages will have more punctuation marks than truthful messages. |
| Number of sentences | Computational Linguistic | Frequency of sentences in messages. Deceptive messages will have more sentences compared to truthful messages |
| Number of adjectives | Computational Linguistic | Deceptive messages can be vague and ambiguous hence adjectives will be used for this purpose |
| Number of adverbs | Computational Linguistic | Deceptive messages are vaguer and more ambiguous hence adverbs will be used for this purpose. Adverbs are words that modify nouns |
| Number of nouns | Computational Linguistic | Frequency of nouns. Deceptive messages will have more nouns compared to truthful messages |
| Number of function words | Computational Linguistic | Frequency of function words. Deceptive messages will have more function words compared to truthful messages |
| I | Psycholinguistic | Deceptive messages contain fewer first personal pronouns, singular *I*, to avoid accountability. |
| Analytic | Psycholinguistic | Deceptive messages will have less analytical words hence low cognitive complexity |
| Six letter words | Psycholinguistic | Words with more than six characters. Measures longer words hence cognitive complexity. Deceptive messages will have fewer words greater than six characters to keep messages less cognitively complex |
| Insight | Psycholinguistic | Deceptive messages will have less insight words hence low cognitive complexity |
| class {0, 1} | | These are the class labels for the binary classifiers. We used a label of 0 for deceptive instances and 1 for truthful instances |

### c) XGBoost Framework and Model Generation

The *eXtreme Gradient Boosting* (XGBoost) machine learning algorithm is a tree ensemble, which uses multiple decision tress called weak learners to solve classification and regression problems. *XGBoost* is a variant of the Gradient Boosting Machine (GBM) algorithm, that is scalable and able to leverage the power of decision tree ensembles for better performance optimization and goodness-of-fit (Chen & Guestrin, 2016). We used an XGBoost binary classifier to train the four hybrid models and evaluate them using a disinformation test set that was not part on the training data. SHAP was then used to evaluate the output of the models to determine the impact of the features in the outcomes.

## 4. Results and Analysis

### a) Evaluating Model Performance

We used these classifier performance metrics to evaluate our *XGBoost* models: precision, recall, F-measure, and Receiver Operating Characteristic (ROC) curves. Precision measures the proportion of the relevant examples in the datasets, which the classifier declared deceptive, while recall measures the number of relevant deceptive examples which were accurately predicted. F-measure is the harmonic mean for precision and recall, and ROC curves illustrate the tradeoff between true positive rate and false positive rate of the classifiers. Table 3 shows the evaluation metrics of the models and generally all the models performed well.

**Table 3: Performance Metrics of the Models**

| Model | Accuracy | Precision | Recall | F1 |
|---|---|---|---|---|
| DIS+EN | 85% | 0.91 | 0.85 | 0.86 |
| DIS+FB | 75% | 0.77 | 0.75 | 0.74 |
| DIS+NEG | 75% | 0.74 | 0.75 | 0.74 |
| DIS+POS | 80% | 0.90 | 0.80 | 0.82 |

As shown in Table 3, the DIS+POS model which was trained on disinformation and favorable fake reviews detected deception and disinformation with 80% predictive accuracy while the DIS+NEG model which is trained on disinformation and negative fake reviews, detected deception and disinformation with 75% predictive accuracy. The DIS+FB model, which is trained on disinformation and scams detected deception and disinformation with 75% accuracy rate. Finally, the DIS+EN model which is trained on disinformation and fraud detected deception and disinformation with 85% predictive accuracy.

### b) Evaluating Model Decisions
#### i) DIS+FB model

The DIS+FB model detected deception and disinformation with 75.00% predictive accuracy. According to Figure 1, the top three features which contributed to the models decision were: *Analytic, Average Word Length and Average Sentence Length.* In this model, the top three impactful features comprised of both computational linguistic and psycholinguistic features. We need to point out the scam dataset was obtained from non-native English speaking cybercriminal gang whose portrayal of deception may be influenced by other native and psychological factors which will require further research.

Furthermore, in Figure 1 the waterfall plot for the DIS+FB model and reveals how each of the features contributes to push the model output from the base value, which is the average model output over the training dataset, to the model output. The red color shows the features pushing the prediction higher while in blue are the features which are pushing the prediction lower. Furthermore, f(x) measures the predicted value of the model given input *x,* which is also called the predicted probability value. For this model, the f(x) is 3.459. E[f(x)] is the expected value of the target variable and it measures the mean of all predictions. For this model the E[f(x)] which is also the base value is 0.032.

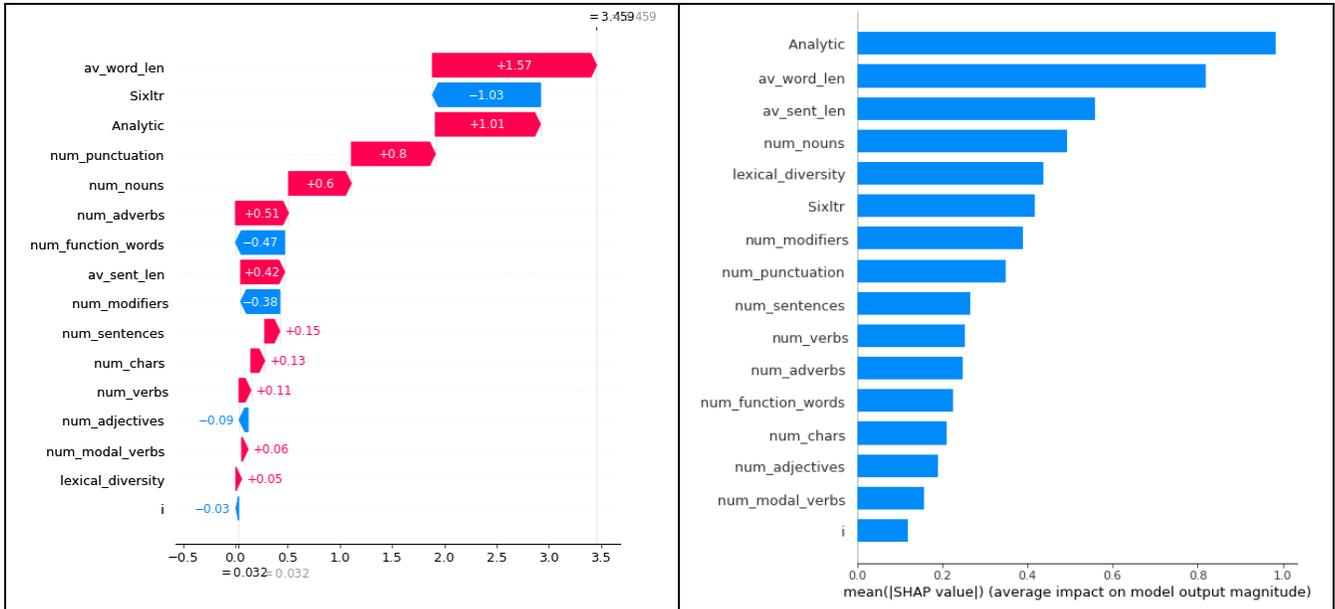

*Figure 1: SHAP Global Summary & Waterfall plot for DIS+FB model*

In this section we are going to discuss the force plots to evaluate which features these the top two features interact with to understand the impact of the relationships.

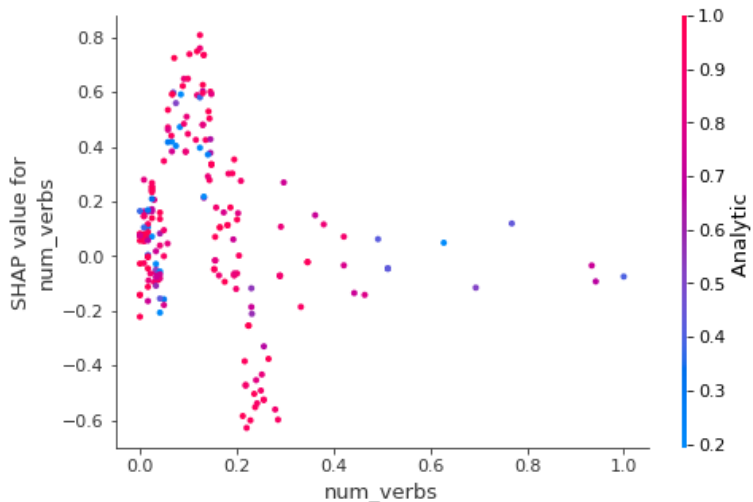

*Figure 2: SHAP Force Plot of Analytic and num_verbs for DIS+FB model*

Figure 2 reveals that there are more verbs to express uncommitted which is a consistent pattern of deception. The relationship of the verb and analytic feature reveals that as fewer verbs were used, there was a strong relationship with analytic words which partly be attributed scams. The scam dataset was obtained from a notorious romance scam native English speaking cybercriminals who use they campaigns to exploit human vulnerabilities of their victims. This is an interesting finding that needs further investigation.

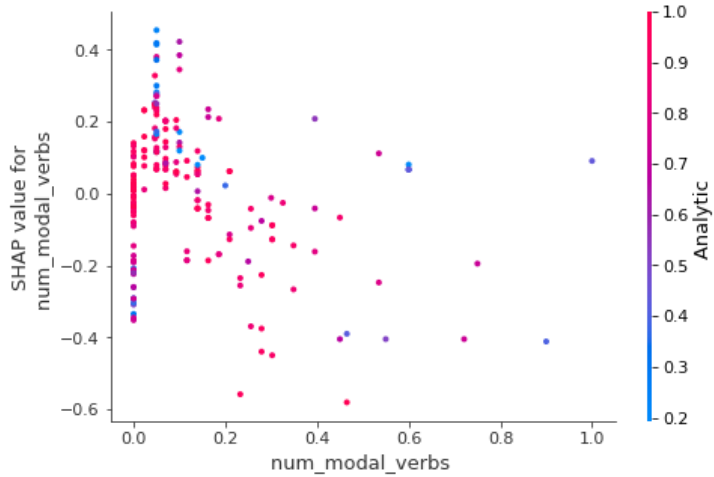

Figure 3: SHAP Force Plot of  num_modal_verbs and Analytic  for DIS+FB model

Figure 3 reveals that there are less modal verbs which is also a consistent deception pattern for expressing non-commitment in communication. Also we observe that this feature has a strong relationship with the Analytic feature.

Figure 4 reveals that generally there are fewer number characters and generally less analytic words. The use of less analytical words is a consistent deception attribute for reducing cognitive complexity of text messages in order to reuse them in campaigns.

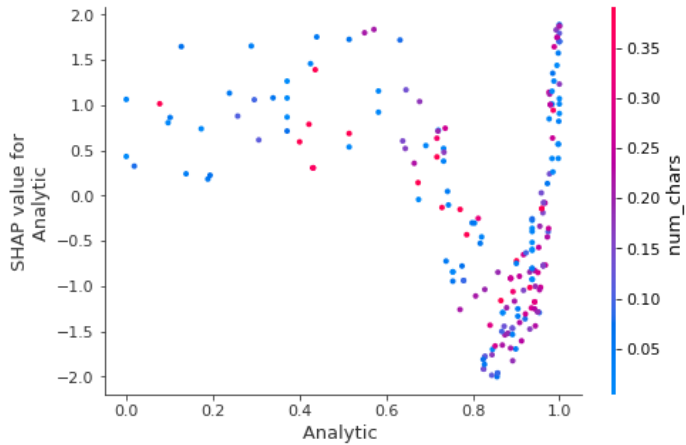

Figure 4: SHAP Force Plot of num_chars and Analytic for DIS+FB model

Figure 5 reveals less use first pronoun I, which is consistent with deception pattern for avoiding accountability of text communication. We also observed that the, I pronoun had a strong relationship with the analytic feature.

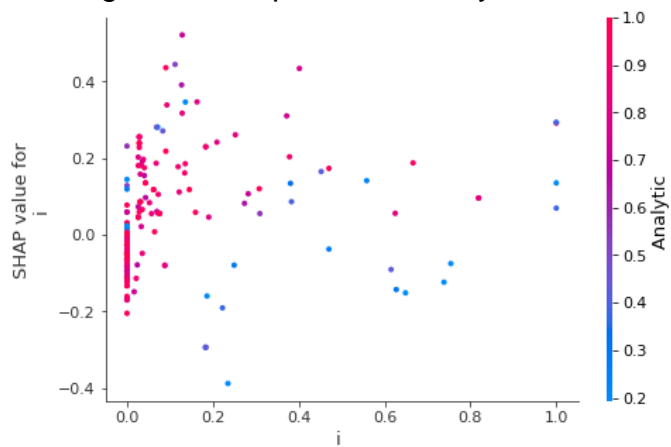

*Figure 5: SHAP Force Plot of Analytic and I for DIS+FB model*

### ii) DIS+EN model

The DIS+EN model detected deception and disinformation with 85.00% predictive accuracy. According to Figure 6, the top three features which contributed to the models decision were: *number of sentences, average sentence length and Analytic.* In this model, the top three impactful features comprised of both computational linguistic and psycholinguistic features. The fraud dataset was obtained from native English speakers from a corporation who were both eloquent in English and also well educated.

Figure 6 shows the waterfall plot for the DIS+EN model and reveals how each of the features contributes to push the model output from the base value, which is the average model output over the training dataset, to the model output. For this model, the f(x) is 3.665. E[f(x)] is the expected value of the target variable and it measures the mean of all predictions. For this model the E[f(x)] which is also the base value is -0.185.

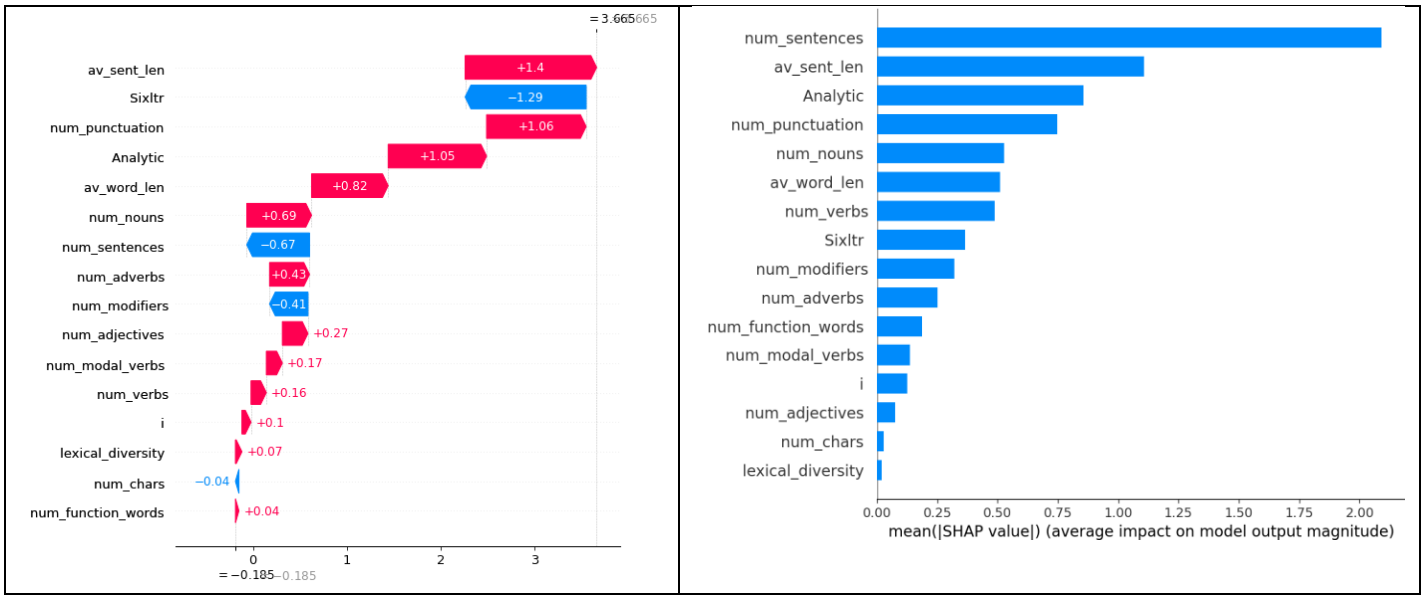

*Figure 6: SHAP Global Summary & Waterfall plot for DIS+EN model*

In this section discuss the force plots to evaluate which features these the top two features interact with to understand the impact of the relationships.

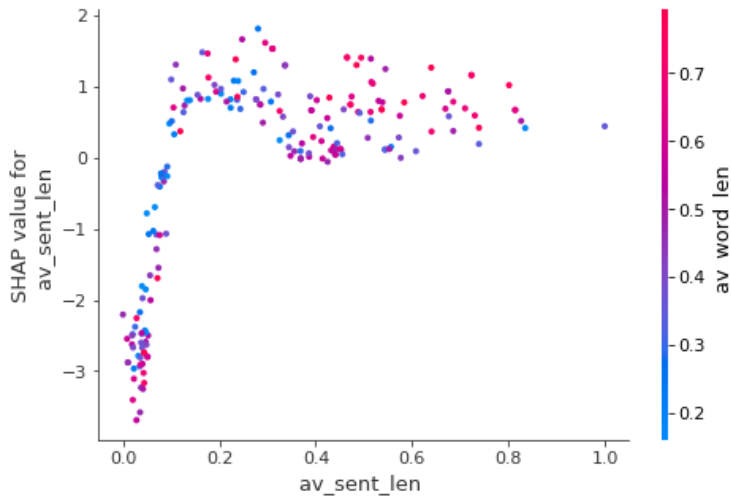

*Figure 7: SHAP Force Plot of av_sent_len and av_word_len for DIS+EN model*

Figure 7 shows that scatterplot average sentence length varies from low to medium high at 0.8. As the spread of the plot increases so that the strength of the average word length. Low average word length reduces cognitive complexity of a deceptive message and makes it more reusable in a campaign.

Figure 8 reveals interaction between number of punctuation and number of sentences. There are fewer number of sentences and limited punctuation which is typical of business communication since the fraud dataset was an email dataset.

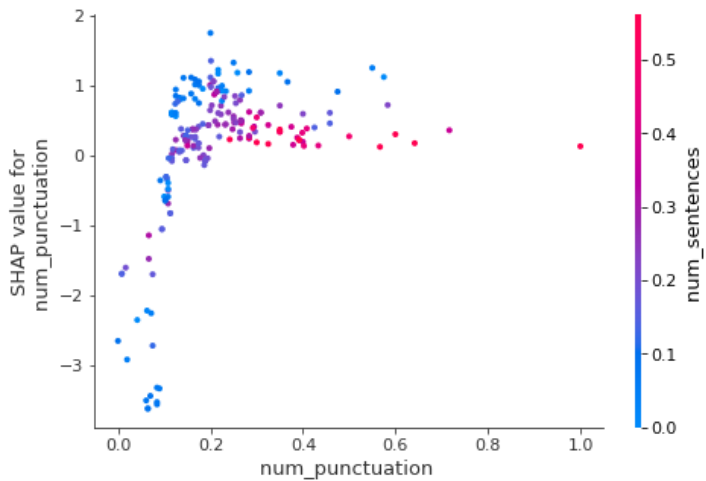

*Figure 8: SHAP Force Plot of num_punctuation and num_sentences for DIS+EN model*

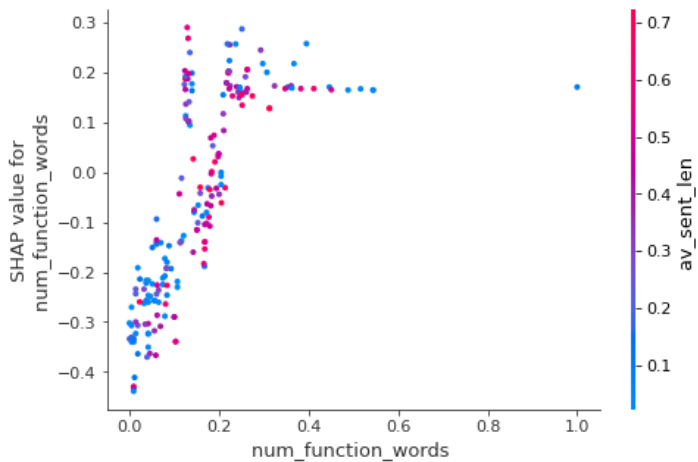

*Figure 9: SHAP Force Plot of num_function_words and av_sent_len for DIS+EN model*

Figure 9 is plot of the interaction between number of function words and average sentence length. The relationship between number of function words and average sentence length is stronger with fewer number of function words.

Figure 10 is plot of the interaction between six-letter words and average sentence length. There is a fair spread of Six-letter words from low to medium and then from medium to high comprising of very low average sentence length, which was reveals that this fraud was orchestrated by an educated people.

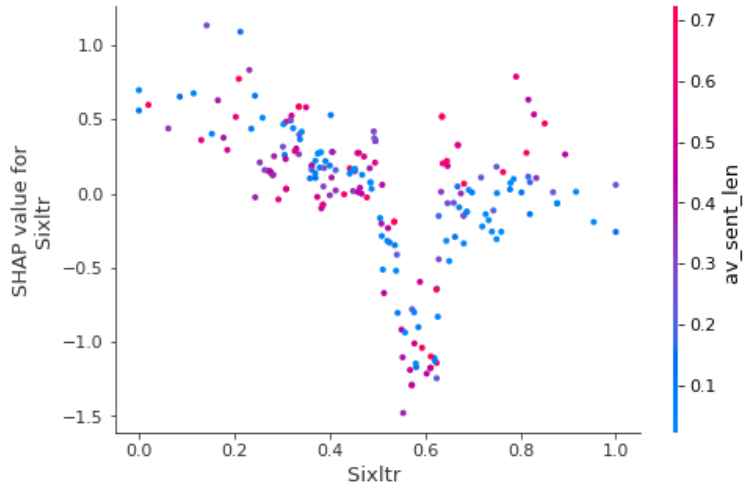

*Figure 10: SHAP Force Plot of Sixltr and av_sent_len for DIS+EN model*

### iii) DIS+NEG model

The DIS+NEG model detected deception and disinformation with 75.00% predictive accuracy. According to Figure 11, the top three features which contributed to the model's decision were: *number of punctuation, number of verbs and Analytic.* In this model, the top three impactful features comprised of both computational linguistic and psycholinguistic features. Figure 11 shows the waterfall plot for the DIS+NEG model and reveals how each of the features contributes to push the model output from the base value, which is the average model output over the training dataset, to the model output. For this model, the f(x) is 3.601. E[f(x)] is the expected value of the target variable and it measures the mean of all predictions. For this model the E[f(x)] which is also the base value is -0.046.

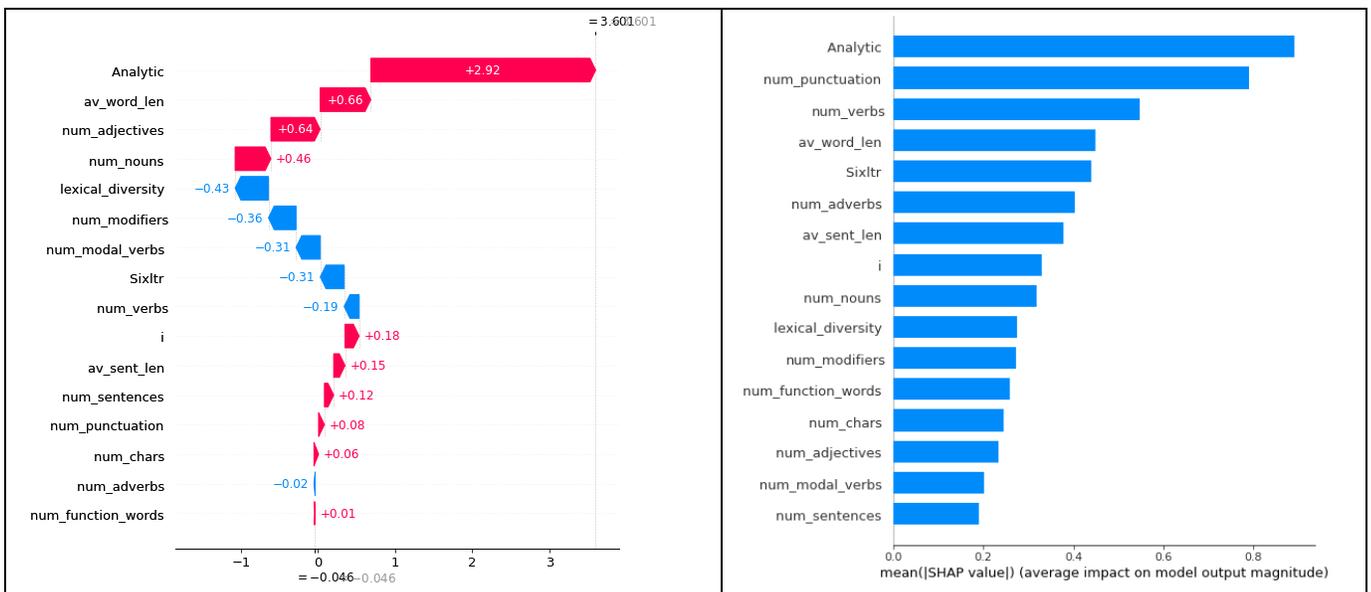

*Figure 11: SHAP Global Summary & Waterfall plot for DIS+NEG model*

In this section we discuss the force plots to evaluate which features these the top two features interact with to understand the impact of the relationships.

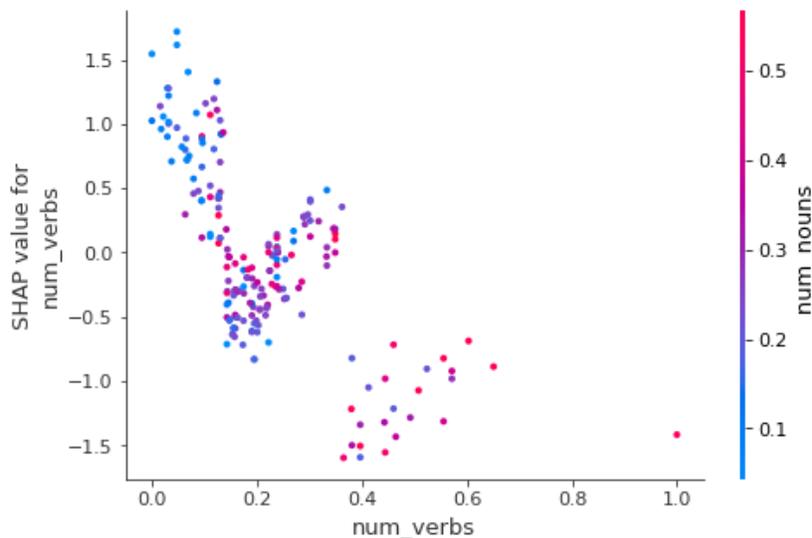

*Figure 12: SHAP Force Plot of num_verbs and num_nouns for DIS+NEG model*

Figure 12 is plot of the interaction between number of verbs as the relationship with number of nouns increases. This contradicts the deception pattern of use of more nouns and verbs to reduce cognitive complexity of text and make it reusable in campaigns.

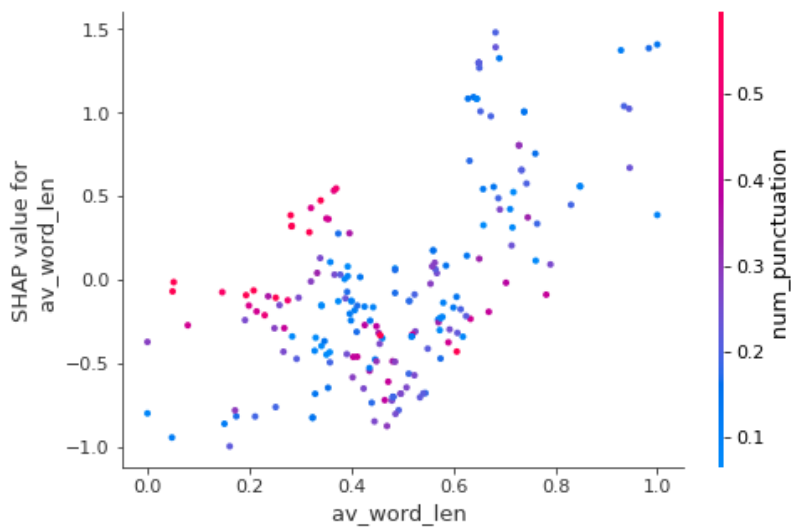

*Figure 13: SHAP Force Plot of av_word_len and num_punctuation for DIS+NEG model*

Figure 13 is plot of the interaction between average word length and number of punctuation marks and in this case as the relationship with punctuation marks reduces, the average word length increases. This is consistent with deception where punctuation marks increase while average word length reduces.

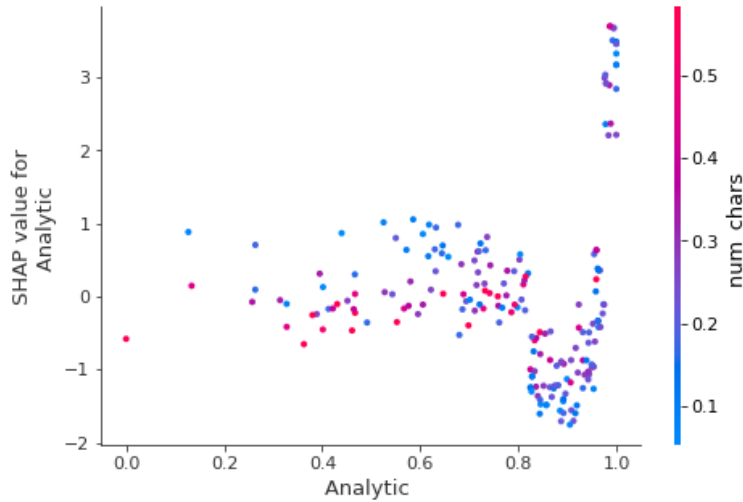

*Figure 14: SHAP Force Plot of Analytical and num_chars for DIS+NEG model*

Figure 14 is plot of the interaction between Analytical and number of characters. The relationship between number of characters and analytical words is average which is consistent with deception.

### iv) DIS+POS model

The DIS+POS model detected deception and disinformation with 80.00% predictive accuracy. According to Figure 15, the top three features which contributed to the model's decision were: *number of punctuation, I and Analytic.* In this model, the top three impactful features comprised of both computational linguistic and psycholinguistic features. Figure 15 shows the waterfall plot for the DIS+POS model and reveals how each of the features contributes to push the model output from the base value, which is the average model output over the training dataset, to the model output. For this model, the f(x) is 3.919. E[f(x)] is the expected value of the target variable and it measures the mean of all predictions. For this model the E[f(x)] which is also the base value is -0.154.

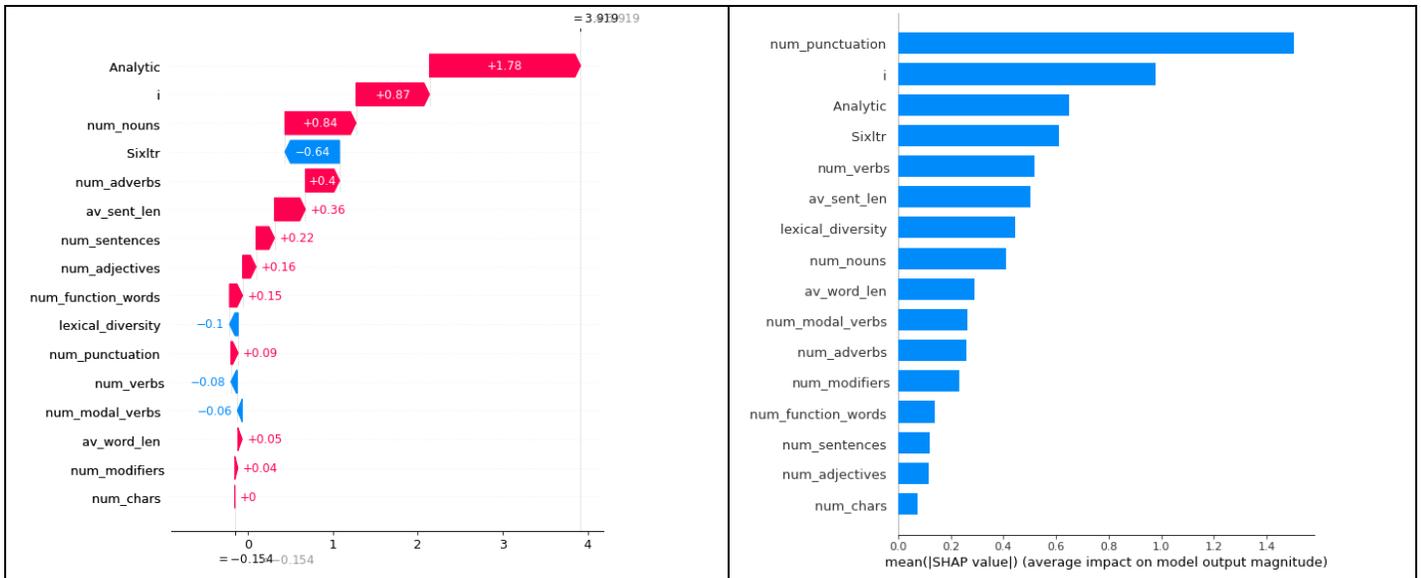
*Figure 15: SHAP Global Summary & Waterfall plot for DIS+POS*

In this section we discuss the force plots to evaluate which features these the top two features interact with to understand the impact of the relationships. Figure 16 is plot of the interaction between average sentence length and I. The scatterplot reveals that there is limited use of self-pronoun I as the average sentence length increases which is also consistent with deception.

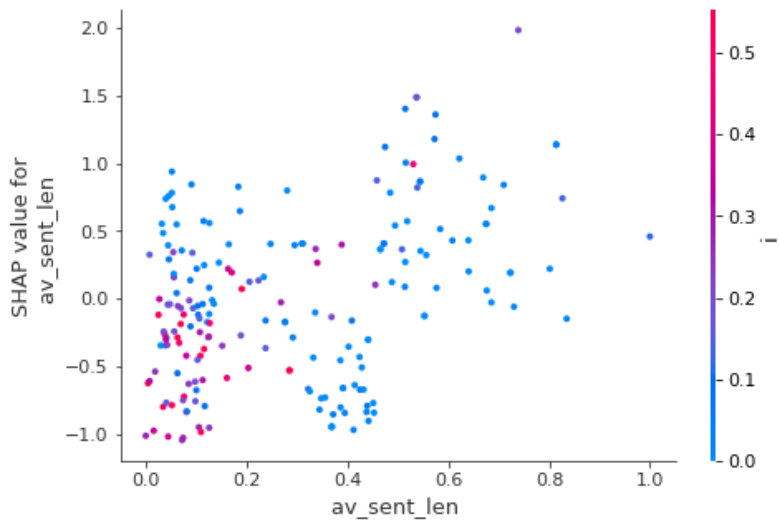
*Figure 16: SHAP Force Plot of average sentence length and I for DIS+POS model*

## 5. Future work and Conclusion

Disinformation continues to be a major problem especially because very sophisticated cybercriminal organizations and nation-states are benefiting from these information and influence operations. This paper demonstrates that it is possible to use explainable machine learning to detect deception and disinformation from trolls using computational linguistic and psycholinguistic processes. Future work will explore patterns of deception in other forms of disinformation campaigns as well as socio-economic factors linked to deception to helps us evaluate generalizability of our models.